\documentclass[aps,prl,twocolumn,floatfix,showpacs,citeautoscript,superscriptaddress,longbibliography,hyperlinks]{revtex4-2}
\usepackage{amsmath}
\usepackage{amssymb}
\usepackage{amsfonts}

\usepackage[breaklinks,hypertexnames=false]{hyperref}
\hypersetup{colorlinks,linkcolor={magenta},citecolor={blue},urlcolor={blue}} 

\usepackage{graphicx}
\usepackage{enumitem}
\setlist[enumerate]{leftmargin=6mm}
\usepackage[caption = false]{subfig}
\usepackage{braket}
\usepackage{siunitx}
\usepackage{tikz}
\usepackage{bbm}
\usepackage{circuitikz}
\usepackage[capitalise]{cleveref}

\usepackage{glossaries}
\glsdisablehyper
\newacronym{ssh}{SSH}{Su-Schrieffer-Heeger}
\newacronym{mbs}{MBS}{Majorana bound state}
\newacronym{abs}{ABS}{Andreev bound state}
\newacronym{car}{CAR}{crossed Andreev reflection}
\newacronym{ec}{EC}{electron co-tunneling}
\newacronym{dos}{LDOS}{local density of states}
\newacronym{gf}{GF}{Green's function}

\newcommand{\md}{\mathrm{d}}

\newcommand{\me}{\mathrm{e}}

\definecolor{mahogany}{RGB}{192,64,0}
\definecolor{magenta2}{RGB}{255,102,255}
\definecolor{crimson}{RGB}{220,20,60}

\usepackage{sidecap,tikz}
\definecolor{lime}{HTML}{A6CE39}
\DeclareRobustCommand{\orcidicon}{\hspace{-1mm}
	\begin{tikzpicture}
		\draw[lime, fill=lime] (0,0) 
		circle [radius=0.16] 
		node[white] {{\fontfamily{qag}\selectfont \tiny \,ID}};
		\draw[white, fill=white] (-0.0525,0.095) 
		circle [radius=0.007];
	\end{tikzpicture}
	\hspace{-3mm}
}
\foreach \x in {A, ..., Z}{\expandafter\xdef\csname orcid\x\endcsname{\noexpand\href{https://orcid.org/\csname orcidauthor\x\endcsname}
		{\noexpand\orcidicon}}
}

\newcommand{\unal}{Departamento de F\'{\i}sica, Universidad Nacional de Colombia, 110911 Bogot\'a, Colombia}
\newcommand{\uam}{Department of Theoretical Condensed Matter Physics\char`,~Universidad Aut\'onoma de Madrid, 28049 Madrid, Spain}
\newcommand{\ifimac}{Condensed Matter Physics Center (IFIMAC), Universidad Aut\'onoma de Madrid, 28049 Madrid, Spain}
\newcommand{\inc}{Instituto Nicol\'as Cabrera, Universidad Aut\'onoma de Madrid, 28049 Madrid, Spain}

\begin{document}

\title{Topological superconductivity in a dimerized Kitaev chain revealed by nonlocal transport}

\author{Rafael Pineda Medina
}
\affiliation{\unal}

\author{Pablo Burset\orcidB{}}
\affiliation{\uam}
\affiliation{\ifimac}
\affiliation{\inc}

\author{William J. Herrera\orcidC{}}
\affiliation{\unal}

\date{\today}

\begin{abstract}
Artificial Kitaev chains engineered from semiconducting quantum dots coupled by superconducting segments offer a promising route to realize and control Majorana bound states for topological quantum computation. We study a dimerized Kitaev chain--equivalent to a superconducting Su-Schrieffer-Heeger model--and analyze the behavior of the resulting two coupled chains. We show that interference between Majorana edge modes from each chain gives rise to observable signatures in nonlocal conductance. Additionally, we identify a parity effect in the system length that governs the coupling of edge states, supported by an analytical model. Our results provide experimentally accessible probes for Majorana hybridization in mesoscopic topological superconductors.
\end{abstract}

\maketitle


\emph{Introduction.}---
\glspl{mbs} in topological superconductors have attracted great interest for their potential application in fault-tolerant quantum computation due to their non-Abelian statistics and inherent protection against local decoherence
~\cite{Nayak2008Sep, Potter2010Nov, GuoHM, 
Lahtinen2017Sep, Lian2018Oct, Prada2020oct}. The Kitaev chain serves as a minimal model for understanding the emergence and manipulation of \glspl{mbs} at the ends of a topological superconductor~\cite{AYuKitaev_2001, Borla2021Jun, Schneider2022Apr, Awoga2024Aug, Frolov2020Jul, Pan2023Jan}. Experimental progress in engineered systems, such as semiconducting nanowires with strong spin-orbit coupling proximized by superconductors, has brought artificial Kitaev chains within reach and enabled the controlled study of topological superconductivity and Majorana physics~\cite{Liu2022Dec,Dvir2023Feb,Bordin2023Sep,TenHaaf2024Jun,Zatelli2024Sep,Tsintzis2024Feb,Bordin2025Mar}. 

Artificial platforms based on arrays of quantum dots coupled via superconducting segments offer enhanced flexibility and tunability~\cite{Pham2022Mar, Kiczynski2022Jun, Lan-Yun2020Apr, Lima2023Jan, Sticlet2014Mar, Hua2019Nov, Kobialka2020Apr, Bahari2021Dec, LeHur2017Nov, Guo2021Feb, Vigliotti2021Jun}. These systems allow for precise control over coupling strengths, chemical potentials, and pairing amplitudes, thereby enabling the exploration of richer phase diagrams and more complex geometries. In particular, the possibility of coupling multiple Kitaev chains introduces new degrees of freedom such as interference between edge modes, parity effects, and nonlocal transport signatures—phenomena with direct implications for braiding and readout schemes in topological qubits~\cite{Acosta2018Jul, McCann2023Jun, Vliet2019}. 

In this work, we theoretically investigate a dimerized Kitaev chain, equivalent to a superconducting \gls{ssh} model, as a tunable platform for probing coupled Majorana physics. The dimerized system can be regarded as two coupled Kitaev chains connected via the onsite energies. We analyze the interplay between topological phases, chain parity, and inter-chain coupling and demonstrate that interference between \glspl{mbs} from different effective chains leads to distinctive nonlocal conductance features. 
In particular, when the chains are coupled and their size is such that the Majoranas at each edge hybridize, the nonlocal conductance displays a characteristic resonance pattern of crossed Andreev processes. 
An analytical model allows us to capture the long range and parity-dependent hybridization of Majorana modes and its role in the nonlocal conductance. 
Our results highlight the utility of dimerized Kitaev systems as building blocks for scalable topological architectures and in the search for unambiguous signatures of Majorana states~\cite{Liu2017Aug, Deng2018Aug, Yavilberg2019Dec, Pan2020Mar, Cayao2021jul, Cayao2021Oct, Pan2021Jun, Lai2022Sep, Feng:2022erg, Marra2022Jan, Aksenov2023Feb, Lu2022Dec, Dutta2024Mar}. 



\begin{figure}[b]
\includegraphics[keepaspectratio=true,width=1.0\linewidth]{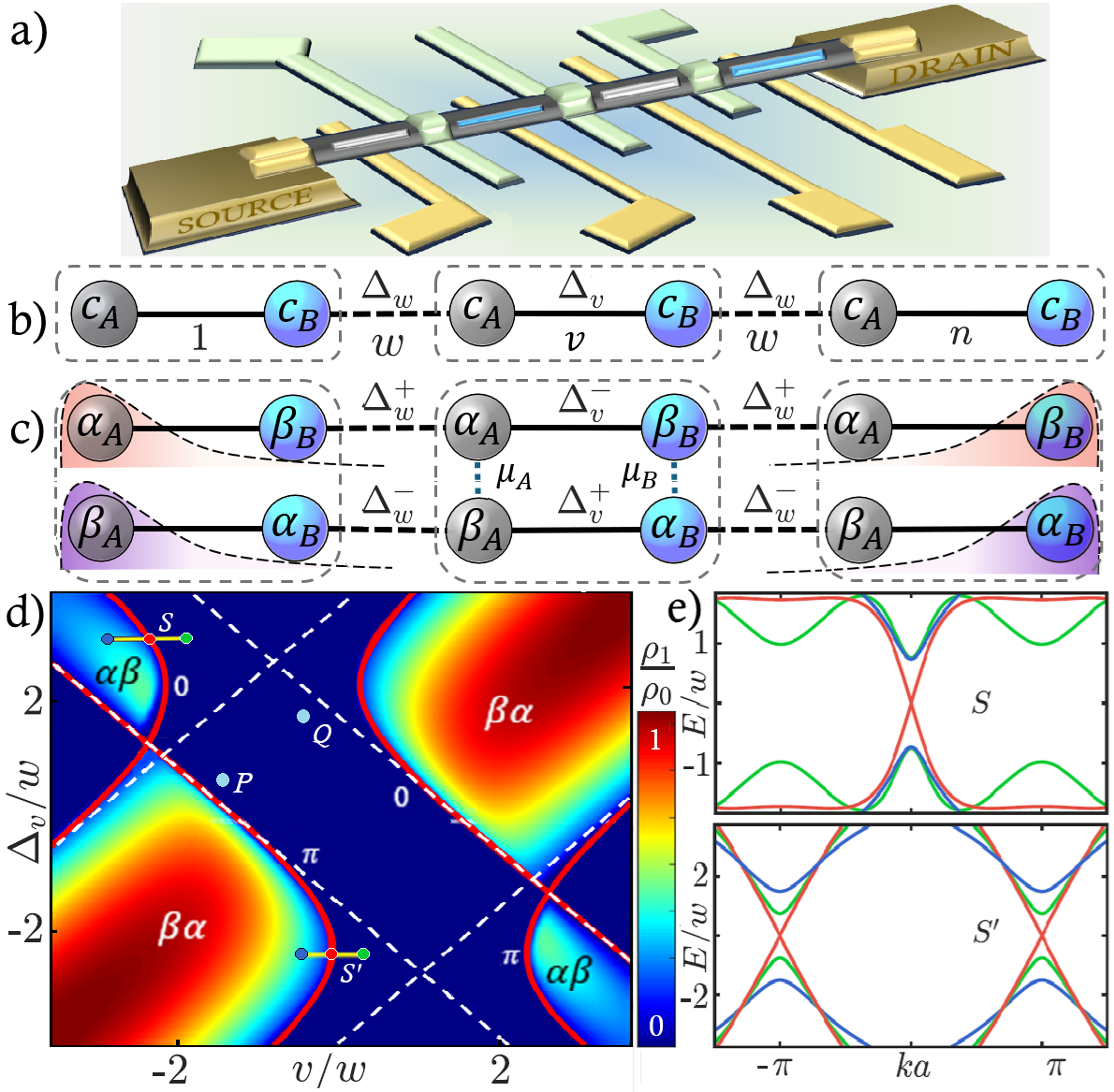}
\caption{Artificial superconducting \gls{ssh} chain. 
    a) Chain of quantum dots (gray and blue regions) coupled by superconducting leads (green). 
    b,c) Dimerized chain in the (b) fermionic and (c) Majorana basis featuring the $\alpha\beta$ and $\beta\alpha$ effective chains. 
    d) $v-\Delta_v$ map of \gls{dos} for $\mu_A=-\mu_B=5w/3$ and $\Delta_{w}=2w$. 
    Red lines delimit nontrivial regions with finite \gls{dos} (white dashed lines for $\mu_{A,B}=0$). 
    e) Bands at the phase transitions denoted as $S$ and $S'$ in d). 
}
\label{fig1}
\end{figure}



\emph{Effective Kitaev chains.}--- 
The Hamiltonian of a dimerized Kitaev chain with two atoms A and B per unit cell of size $a$ [\cref{fig1}b)], is
\begin{align}
    H ={}& \sum_{\tau} \sum_{j=1}^n\left( \mu_{\tau}c^{\dagger}_{\tau j}c_{\tau j}+vc^{\dagger}_{\bar{\tau} j}c_{\tau j}+\Delta_{v}c^{\dagger}_{ \tau j}c^{\dagger}_{\bar{\tau} j} \right) 
    \notag \\
    + & \sum_{\tau}\sum_{j=1}^{n-1} \left( wc^{\dagger}_{\tau j+i}c_{ \bar{\tau} j} +\Delta_{w}c^{\dagger}_{\tau j}c^{\dagger}_{j+1 \bar{\tau j}} \right)+ \text{H.c.} , 
 \label{eqH}
\end{align}
where $c^{\dagger}_{\tau j}$ ($c_{\tau j}$) are creation (annihilation) fermionic operators on atom $\tau=(A,B)$ of site $j$, and we consider, for simplicity, that all atoms of the same type have the same onsite potential $\mu_\tau$. 
Normal and superconducting intra-cell (inter-cell) hoppings between A and B atoms are respectively denoted by $v$ and $\Delta_v$ ($w$ and $\Delta_w$). 
The Hamiltonian in \cref{eqH} can be decomposed into two Kitaev effective chains by changing into a Majorana representation where fermion operators are written as $c_{\tau j} = (\alpha_{\tau j}+i\beta_{\tau j})/\sqrt{2}$, with $\alpha_{\tau j}=\alpha_{\tau j}^\dagger$ and $\beta_{\tau j}=\beta_{\tau j}^\dagger$ self-adjoint operators such that
\begin{equation}\label{eq:commutation}
\{ \alpha_{\tau j},\alpha_{\tau' k}\}=2\delta_{\tau\tau',jk}~, \; 
\{ \beta_{\tau j},\beta_{\tau' k}\}=2\delta_{\tau\tau',jk}~,  
\end{equation}
and $\{ \alpha_{\tau j},\beta_{\tau' k}\}=0$. In this Majorana representation \cref{eqH} takes the form
\begin{equation}\label{eq:Hmajo}
 H= -\frac{i}{2} \left( h_{\alpha\beta} + h_{\beta\alpha} \right) + \frac{1}{2} \sum_{\tau,j} \mu_{\tau} (1 + i \alpha_{\tau j}\beta_{\tau j}) ,
\end{equation}
with
\begin{subequations}\label{eq:Hab}
\begin{align}
h_{\alpha\beta} =& \sum_{j=1}^{n} \Delta^{-}_{v} \alpha_{A j}\beta_{Bj}
    + \sum_{j=1}^{n-1} \Delta^{+}_{w} \beta_{Bj}\alpha_{Aj+1} , \\
h_{\beta\alpha} =& \sum_{j=1}^{n} \Delta^{+}_{v} \beta_{A j}\alpha_{Bj}
    + \sum_{j=1}^{n-1} \Delta^{-}_{w} \alpha_{B j}\beta_{A j+1} ,
\end{align}
\end{subequations}
where $\Delta^{\pm}_{v}=\Delta_{v}\pm v$ and $\Delta^{\pm}_{w}=\Delta_{w}\pm w$ respectively are the intra-cell and inter-cell effective couplings. 
The dimerized Kitaev chain is thus decomposed into two independent Majorana chains that alternate between $\alpha_\tau$ and $\beta_{\bar{\tau}}$ type of Majorana states, see \cref{fig1}c). Each chain is labeled according to the Majoranas of the atoms from the first cell, $\alpha\beta$ and $\beta\alpha$. 
Importantly, the coupling between these chains depends only on the local onsite energies $\mu_{A}$ and $\mu_{B}$, and the chains are thus completely decoupled for $\mu_{A}=\mu_{B}=0$. 

%
%
%
The $\mathbb{Z}_{2} $ invariant associated to \cref{eqH} is computed from the skew-symmetric matrix $H_{s}(k)$, obtained after rotation of the bulk Hamiltonian, see Supplemental Material~\cite{SM} for more details:
\begin{equation}\label{eq:invariant}
(-1)^\nu= \prod_{k=0,\pi}\mathrm{sgn} \left(\mathrm{Pf}[H_s(k)] \right)= 
\mathrm{sgn}(Q_{0})\mathrm{sgn}(Q_{\pi}),
\end{equation}
with $\mathrm{Pf}[A]$ the Pfaffian of matrix $A$, 
\begin{equation}\label{eq:Qs}
Q_{\eta} =(\Delta^{+}_{v}+\eta\Delta^{-}_{w})(\Delta^{-}_{v}+\eta\Delta^{+}_{w})+\mu_{A}\mu_{B},
\end{equation}
and $\eta=\pm1$ for $k=0,\pi$, respectively. 

When the chains are decoupled, $\mu_{A}=\mu_{B}=0$, the $\alpha\beta$ ($\beta\alpha$) chain becomes topologically nontrivial for 
$ |\Delta^{+}_{w}| > |\Delta^{-}_{v}| $ ($ |\Delta_{w}^{-}|>|\Delta_{v}^{+}|$), in analogy to the \gls{ssh} model~\cite{PINEDAMEDINA2025115729}. 
When both conditions are satisfied, the system exhibits a trivial phase. 
The topological regions for the decoupled chains are indicated by white dashed lines in \cref{fig1}d). 

By contrast, when we consider finite onsite potentials $\mu_{A,B}\neq0$, the effective chains couple and the phase diagram becomes more complex. For example, for $\mu_{A}=-\mu_{B}$ the phase transition lines become hyperbolas [solid red lines in \cref{fig1}d)]. Across these lines, \cref{eq:invariant} changes sign indicating a closing and re-opening of the bulk band gap, as shown in \cref{fig1}e) for two examples denoted as $S$ and $S'$ in \cref{fig1}d). 
We further verify the bulk-boundary correspondence by computing the \gls{dos} at the edge of a semi-infinite chain, $\rho_{1}=-\mathrm{Im}[\mathrm{Tr}(\hat{G}^{11}_{AA})]/\pi$, with $\hat{G}^{jk}_{\tau\tau'}$ the \gls{gf} associated to the Hamiltonian in \cref{eqH}~\cite{SM}. 
The zero-energy \gls{dos} develops peaks at the topologically nontrivial regions corresponding to $\alpha\beta$ and $\beta\alpha$ chains, as shown in \cref{fig1}d). 

\begin{figure}[t!]
\includegraphics[keepaspectratio=true,width=1.0\linewidth]{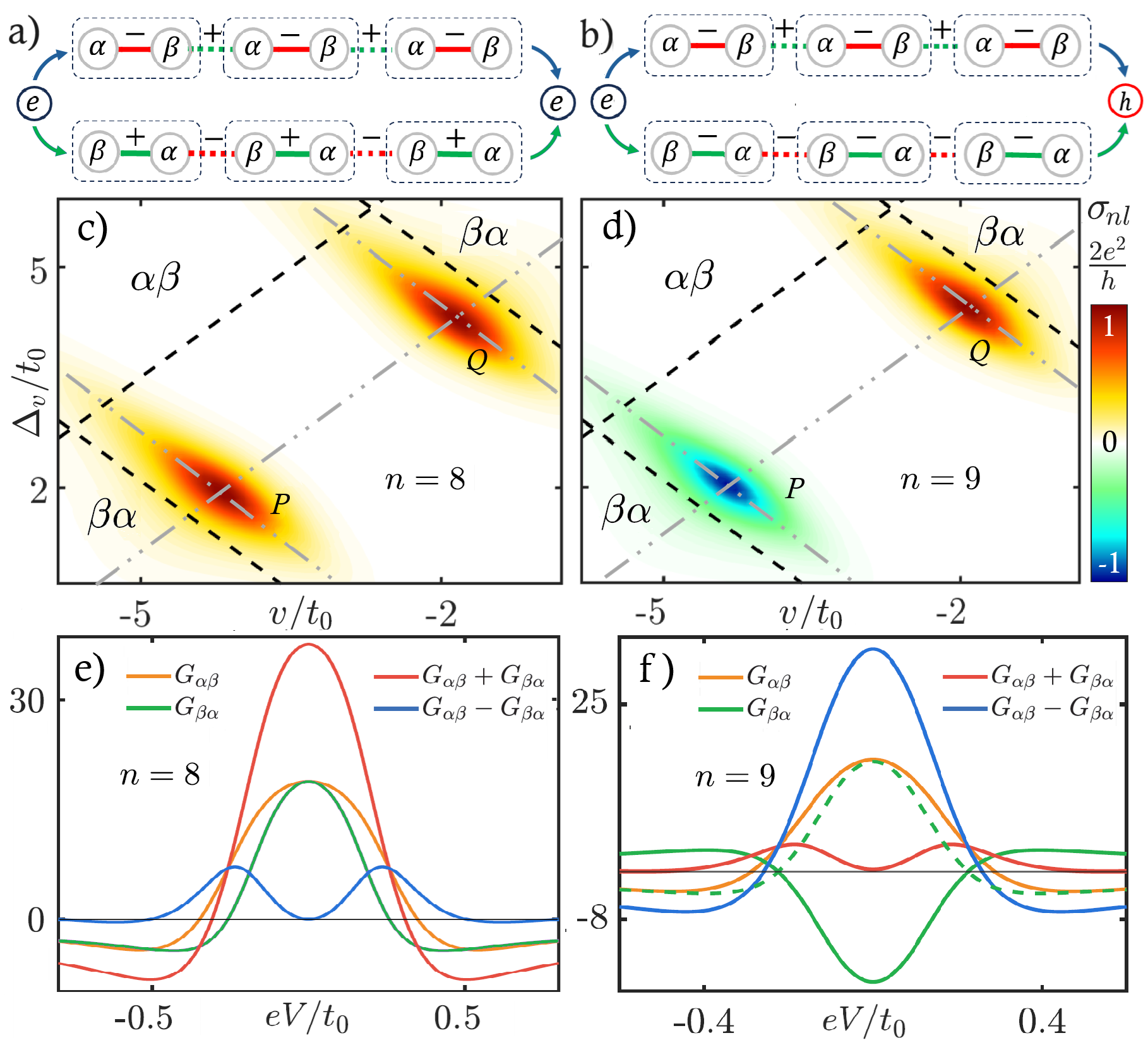}
\caption{Finite-size effects. 
a,b) Sketch of the configuration of effective Majorana chains leading to an \gls{ec} (a) or a \gls{car} (b) process. 
Green and red solid (dashed) lines respectively indicate $\Delta_v^+$ ($\Delta_w^+$) and $\Delta_v^-$ ($\Delta_w^-$) hoppings, with sign shown on top of each line. 
c,d) Zero-bias nonlocal conductance for chains with $\mu_A=\mu_B=0$ and $n=8$ (c), $n=9$ (d). Gray dot-dashed lines mark the boundaries of topological regions (dashed black lines for the bulk system). 
e,f) Voltage dependence of the Majorana nonlocal correlators $G_{\epsilon\epsilon'}$ at point $P$ for $n=8$ (e) and $n=9$ (f). The dashed line in f) corresponds to $G_{\beta\alpha}$ at point $Q$. 
In all cases, $w=0.3$ and $\Delta_w=0.6$. 
    }
\label{fig2}
\end{figure}

\emph{Majorana interference effect in nonlocal conductance.}---
We now focus on the interesting situation where the two effective chains couple and how the interplay between the \glspl{mbs} at the edges of each chain impact observables like the nonlocal conductance. 
To analyze the electric transport, we couple two metallic electrodes at the edges of a finite chain of $n$ cells and define the nonlocal differential conductance~\cite{SM}
\begin{equation}\label{eq:nonloc}
\sigma_{nl}=\frac{2e^{2}}{h}(T_\text{CAR}-T_\text{EC}), 
\end{equation}
with $e$ the electron charge, $h$ Planck's constant and
\begin{equation}\label{eq:car-ec}
T_\text{CAR}= t_0^4 \rho^{L}_{e}\rho^{R}_{h} |\hat{G}^{1n}_{AB,eh}|^2 , \,\,
T_\text{EC} = t_0^4 \rho^{L}_{e}\rho^{R}_{e} |\hat{G}^{1n}_{AB,ee}|^2 .
\end{equation}
Here, $\rho_{e(h)}^X$ is the \gls{dos} for electron-like (hole-like) quasiparticles on metallic lead $X=L,R$. For simplicity, we only consider a symmetric configuration where the leads are coupled by hopping terms $t_L=t_R\equiv t_0=0.1t$, with $t$ the hopping of the metallic lead. 

We focus on the conductance on electrode $R$, so that $T_\text{CAR}$ and $T_\text{EC}$ are, respectively, the \gls{car} and \gls{ec} transmission probabilities for an electron from $L$ [\cref{fig2}a,b)]. 
The \glspl{gf} in \cref{eq:car-ec} are expressed in the Nambu (fermionic) basis, but we can determine the contribution of each effective chain by changing into the Majorana representation: 
\begin{align} \label{eq:gfgen}
\hat{G}^{jk}_{AB,ee}= {}& \hat{G}^{jk}_{AB,\alpha \beta}-\hat{G}^{jk}_{AB,\alpha  \alpha}-\hat{G}^{jk}_{AB,\beta \alpha}+\hat{G}^{jk}_{ AB,\beta \beta}, \\
\hat{G}^{jk}_{AB,eh}={}& \hat{G}^{jk}_{AB, \alpha \beta } - \hat{G}^{jk}_{AB,\alpha \alpha } + \hat{G}^{jk}_{AB, \beta \alpha} - \hat{G}^{jk}_{AB,\beta \beta}.
\end{align}
Since we are only interested in the nonlocal \glspl{gf} between the edges of the dimerized Kitaev chain, we define the short-hand notation $\hat{G}^{1n}_{AB,\epsilon \epsilon^{\prime}} \equiv G_{\epsilon \epsilon^{\prime}}$, for $\epsilon, \epsilon^{\prime} = \alpha, \beta$. 
The function $G_{\alpha\beta(\beta\alpha)}$ is then the propagator of Majorana states for the $\alpha\beta$ ($\beta\alpha)$ effective chain from the leftmost, A-type atom on cell 1 to the rightmost, B-type atom on cell \textit{n} at the opposite edge of the chain. 
Analogously, $G_{\alpha\alpha(\beta\beta)}$ are the propagators \textit{between} effective chains. 

As a result, the nonlocal conductance in \cref{eq:nonloc,eq:car-ec} describes an interference between the Majorana correlators from the $\alpha\beta$ and $\beta\alpha$ effective chains. 
For example, choosing $\mu_{A} = p\mu_{B}$, with $p=\pm 1$, the effective chains couple through the correlators $G_{\alpha \alpha }=p G_{\beta   \beta }$, and the nonlocal probabilities reduce to the simple expressions
\begin{equation} \label{eq:car-ec-red}
T_\text{CAR(EC)}=t_0^4\rho^{L}_{e}\rho^{R}_{h(e)}|G_{\alpha \beta}\pm G_{\beta \alpha}-(1\pm p)G_{\alpha \alpha}|^{2}.
\end{equation}

To explore the interference of Majorana chains implicit in \cref{eq:car-ec-red} we first focus on the case with decoupled chains, i.e., for $\mu_A=\mu_B=0$. 
\Cref{fig2}c) shows a $v$-$\Delta_v$ map of the nonlocal conductance at zero bias for a finite chain with an even number of cells ($n=8$). The nonlocal conductance is only nonzero at the crossings of the topological regions corresponding to the $\alpha \beta$ and $\beta \alpha$ chains (gray dot-dashed lines), denoted $P$ and $Q$. Note that the finite chain size has changed the boundaries of the topological regions, reducing their areas as compared to the white dashed lines of \Cref{fig1}d) for a bulk system. At crossing points $P$ and $Q$ the nonlocal conductance is dominated by \gls{car} processes ($\sigma_{nl}=2e^2/h$). 
By decoupling the effective chains, we have that $G_{\alpha\alpha}=0$ and, consequently, $T_\text{CAR,EC}\propto|G_{\alpha\beta}\pm G_{\beta\alpha}|^{2}$. Inside the topological region for the $\alpha \beta$ ($\beta \alpha$) chain, we further have that $G_{\alpha\beta}\neq0$ while  $G_{\beta\alpha}=0$ ($G_{\alpha\beta}=0$, $G_{\beta\alpha}\neq0$), yielding that $T_\text{CAR}= T_\text{EC}$ and $\sigma_{nl}=0$. 

By contrast, at the crossing points $P$ and $Q$ in \cref{fig2}c-d), the $\alpha\beta$ and $\beta\alpha$ chains can interfere due to a finite-size effect, leading to constructive or destructive interference patterns. 
As a result, $\sigma_{nl}$ can exhibit quantized maxima or minima $\sigma_{nl}=\zeta2e^2/h$, with 
\begin{equation}\label{eq:invariant-2}
    \zeta=\left[ \mathrm{sgn}(\Delta^{+}_{w}\Delta^{-}_{w}) \right]^{n-1} 
    \left[ \mathrm{sgn}(\Delta^{+}_{v}\Delta^{-}_{v}) \right]^{n}. 
\end{equation}
Constructive interference between the chains results in $T_\text{CAR}\propto 4|G_{\alpha\beta}|^{2}$ and $T_\text{EC}=0$ [point $Q$, \cref{fig2}d)], whereas destructive interference leads to $T_\text{CAR}=0$ and $T_\text{EC}\propto 4|G_{\alpha\beta}|^{2}$ [point $P$, \cref{fig2}d)]. 
The Majorana correlators $G_{\epsilon\epsilon'}$ are real and, for a fixed chain length, e.g., $n=9$ in \cref{fig2}d), the sign of the correlator $G_{\alpha \beta}$ is given by $\mathrm{sgn}[(\Delta^{+}_{w})^{n-1}(\Delta^{-}_{v})^{n}]$, while the sign of $G_{\beta\alpha}$ is $\mathrm{sgn}[(\Delta^{-}_{w})^{n-1}(\Delta^{+}_{v})^{n}]$, see \cref{fig2}a) and b). 
Importantly, constructive interference occurs when the sign of the propagators $G_{\alpha\beta}$ and $G_{\beta\alpha}$ is the same [\cref{fig2}e)], and when they have opposite signs we find destructive interference [\cref{fig2}f)]. 
Thus, the relative sign between the \gls{ec} and \gls{car} processes is determined by \cref{eq:invariant-2} and depends on the parity of the chain length or number of cells $n$. 
For a chain with an odd number of cells, the interference depends on the relative sign of intracell hoppings $\Delta^{\pm}_{v}$. 
By contrast, for an even-length chain, it depends on the sign of the intercell couplings $\Delta^{\pm}_{w}$.

\begin{figure}[t!]
\includegraphics[keepaspectratio=true,width=0.9\linewidth]{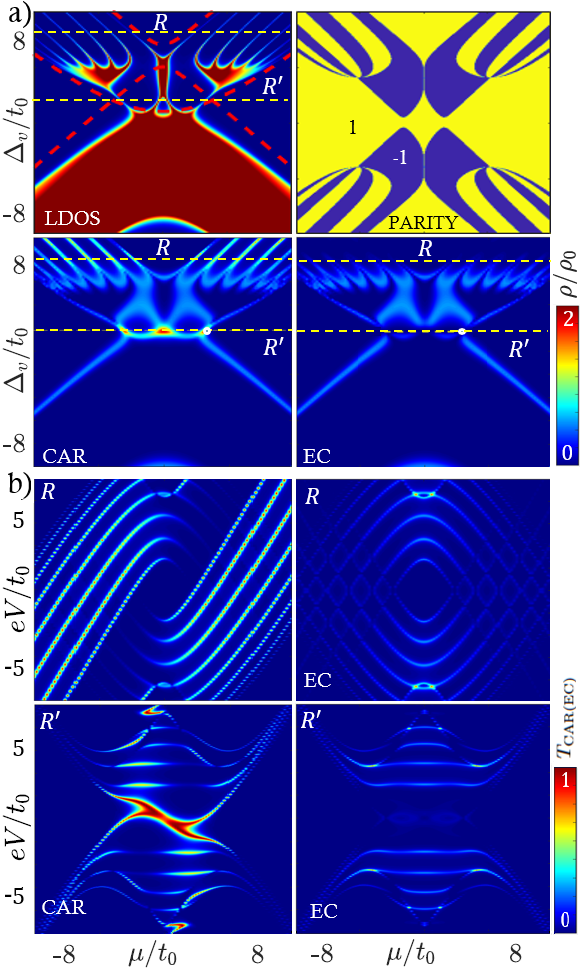}
\caption{ Nonlocal transport for coupled chains. 
a) Zero-bias \gls{dos} (top left), finite-size fermion parity (top right), and \gls{car} and \gls{ec} probabilities (bottom) as a function of $\Delta_v$ and $\mu\equiv\mu_A=-\mu_B$. The dot-dashed red line marks the bulk topological regions. 
b) $T_\text{CAR,EC}$ as a function of the voltage $eV$ and $\mu$ for the $R$ (top) and $R'$ (bottom) lines in a). 
c) \Gls{dos} for electrons and holes at the leftmost A-type and rightmost B-type sites. The arrows indicate the combinations responsible for the \gls{car} and \gls{ec} contributions. 
All panels in a,b,c) are computed for $n=8$ with $v=-0.4$,$w=0.3$, $\Delta_{w}=0.6$ and $t_{0}=0.1$.  
}
\label{fig3}
\end{figure}

\emph{Oscillations of \glspl{mbs} and the role of onsite energies.}---
Having studied the finite-size effects on decoupled effective chains, we can now set $\mu_{A,B}\neq0$ and consider the coupling between $\alpha\beta$ and $\beta\alpha$ chains. 
The correlators $G_{\alpha\alpha}$ and $G_{\beta\beta}$ are now finite, and the behavior of $T_\text{CAR}$ and $T_\text{EC}$ according to \cref{eq:car-ec-red} becomes richer. 
For simplicity, we analyze the case with $\mu\equiv \mu_{A}=-\mu_{B}$ which yields $G_{\beta\beta}=-G_{\alpha\alpha}$ (we examine the instance $\mu_{A}=\mu_{B}$ in the End Matter). 

We plot the zero-energy \gls{dos} at the edge of a finite chain with $n=8$ in \cref{fig3}a), and show with dashed red lines the topological phase diagram corresponding to a bulk system with the same parameters. 
As explained above for the decoupled system, the finite-size effect results in the hybridization of the edge states with two important consequences: 
First, the critical lines in the bulk phase diagram do not exactly coincide with the zero-energy states from the \gls{dos} 
[the \gls{dos} peaks for the semi-infinite chain featured in \cref{fig1}d) perfectly match the topological phase diagram]. 
Also, as we change $\mu$ in \cref{fig3}a), the \gls{dos} in some regions (above line $R'$) exhibits resonant peaks indicating the presence of \glspl{mbs} only on-resonance. 
The number of these resonance peaks is determined by the chain length $n$ and they appear in regions where the chain fermion parity changes, [top-right panel of \cref{fig3}a)]~\cite{SM}. 
The fermionic parity of the system undergoes sign changes as the chemical potential varies, leading to Majorana oscillations. Indeed, a zero-energy state emerges precisely at the parity transition, as shown in the top panels \cref{fig3}a). 
As a result, $T_\text{CAR}$ becomes dominant at these parity flips [bottom panels of \cref{fig3}a)], with resonances at $eV = 0$ corresponding to the spikes observed in the \gls{dos} which dominate the nonlocal conductance. 

We can analyze in more detail the Majorana edge states forming on resonance. 
Expanding the continuous Hamiltonian in \cref{eq:Hmajo} near the band inversion points $k=0,\pm\pi$, along line $R$, we can approximate the spatial dependence of the \glspl{mbs} wavefunction as $\psi(x) \propto \exp(\pm x/\xi\pm i\kappa x)$, where $\xi$ determines the decay length and $\kappa$ characterizes the spatial oscillation. 
The decay length along line $R$ diverges for both the $\alpha\beta$ and $\beta\alpha$ chains, so the \glspl{mbs} become extended oscillatory modes with wavenumber $\kappa^{0,\pi}\propto \sqrt{Q_{0,\pi}}$, see more details in End Matter. The \gls{mbs} oscillations are thus related to the topological invariant in \cref{eq:invariant}, determine the overlap between edge states, and yield the resonant behavior of the nonlocal transmission. 

In general, the formation of the nonlocal transport channels depends on the interplay between the chain length $n a$ and the \glspl{mbs} localization length $\xi$, which is determined by the system parameters. 
As we change $\Delta_v$ in \cref{fig3}a), we can distinguish a region below line $R'$ where the localization length is very small, $\xi\to a$, so that the nonlocal transport is negligible. 
Above line $R'$ the localization length becomes comparable to the chain size, $\xi\gtrsim n a$, and we observe strong, \gls{car}-dominated nonlocal transport on resonance. 
The \glspl{mbs} localization length diverges ($\xi/a\to\infty$) along line $R$, which fulfills the condition $\Delta_v^+/\Delta_w^-=\Delta_v^-/\Delta_w^+$ corresponding to the transition from the $\alpha\beta$ to the $\beta\alpha$ effective chains~\cite{SM}. 

The connection between hybridized states and nonlocal transport yields that the dominant \gls{car} processes are not restricted to zero energy. We thus explore in \cref{fig3}b) the voltage dependence of $T_\text{CAR,EC}$ along the lines $R$ (top panels) and $R'$ (bottom panels). 
The sequence of resonances on line $R$ and the peaks at low $\mu$ for $R'$ are maintained at finite bias with dominant \gls{car} probability. In both cases the resulting nonlocal conductance is very asymmetric with the bias. 
We can readily understand why \gls{car} processes are dominant by comparing the electron \gls{dos} at the leftmost A-type site of the chain ($\rho^L_{Ae}$) to both the electron and hole \gls{dos} on the rightmost B-type site ($\rho^R_{Be,h}$), see \cref{fig3}c) and \cref{eq:car-ec}. 
Since the edge sites are of different type, their onsite energies have opposite signs. 
As a result, $\rho^{L}_{Ae}$ is always on resonance with $\rho^{R}_{Bh}$ but not with $\rho^{R}_{Be}$, as indicated by the arrows in \cref{fig3}c); thus resulting in an energy-filtering mechanism that enhances \gls{car} processes~\cite{Brinkman2010,Pablo_NT_PhysRevB}. 

%


\emph{Conclusions.}---
We demonstrated that the nonlocal conductance of a dimerized Kitaev chain arises from interference between Majorana modes belonging to two different effective Kitaev chains. 
The onsite energy $\mu_{A,B}$ controls the coupling between these effective chains, which can be completely decoupled for $\mu_{A,B}=0$. 
At finite onsite energy, finite-length chains exhibit hybridization of edge states, with both effective chains supporting Majorana modes. This results in the interference between the two Majorana modes, producing conductance signatures that depend on chain parity. 
These interference effects give rise to multiple conductance peaks that scale with chain length and correlate with fermion parity transitions. Additionally, Majorana modes may decay slowly under some conditions and exhibit spatial oscillations along the chain. 
 
R.P.M. and W.J.H. acknowledge support from project ``Ampliación del uso de la mecánica cuántica desde el punto de vista experimental y su relación con la teoría, generando desarrollos en tecnologías cuánticas útiles para metrología y computación cuántica a nivel nacional", BPIN 2022000100133 from SGR of MINCIENCIAS; Gobierno de Colombia. P.B. acknowledges support by Spanish CM ``Talento Program'' project No.~2019-T1/IND-14088 and No.~2023-5A/IND-28927, the Agencia Estatal de Investigaci\'on project No.~PID2020-117992GA-I00, No.~PID2024-157821NB-I00 and No.~CNS2022-135950 and through the ``María de Maeztu'' Programme for Units of Excellence in R\&D (CEX2023-001316-M). 


\nocite{*}
\bibliography{biblio}


\clearpage
\section{End matter}

\begin{figure}[h]
\includegraphics[keepaspectratio=true,width=0.9\linewidth]{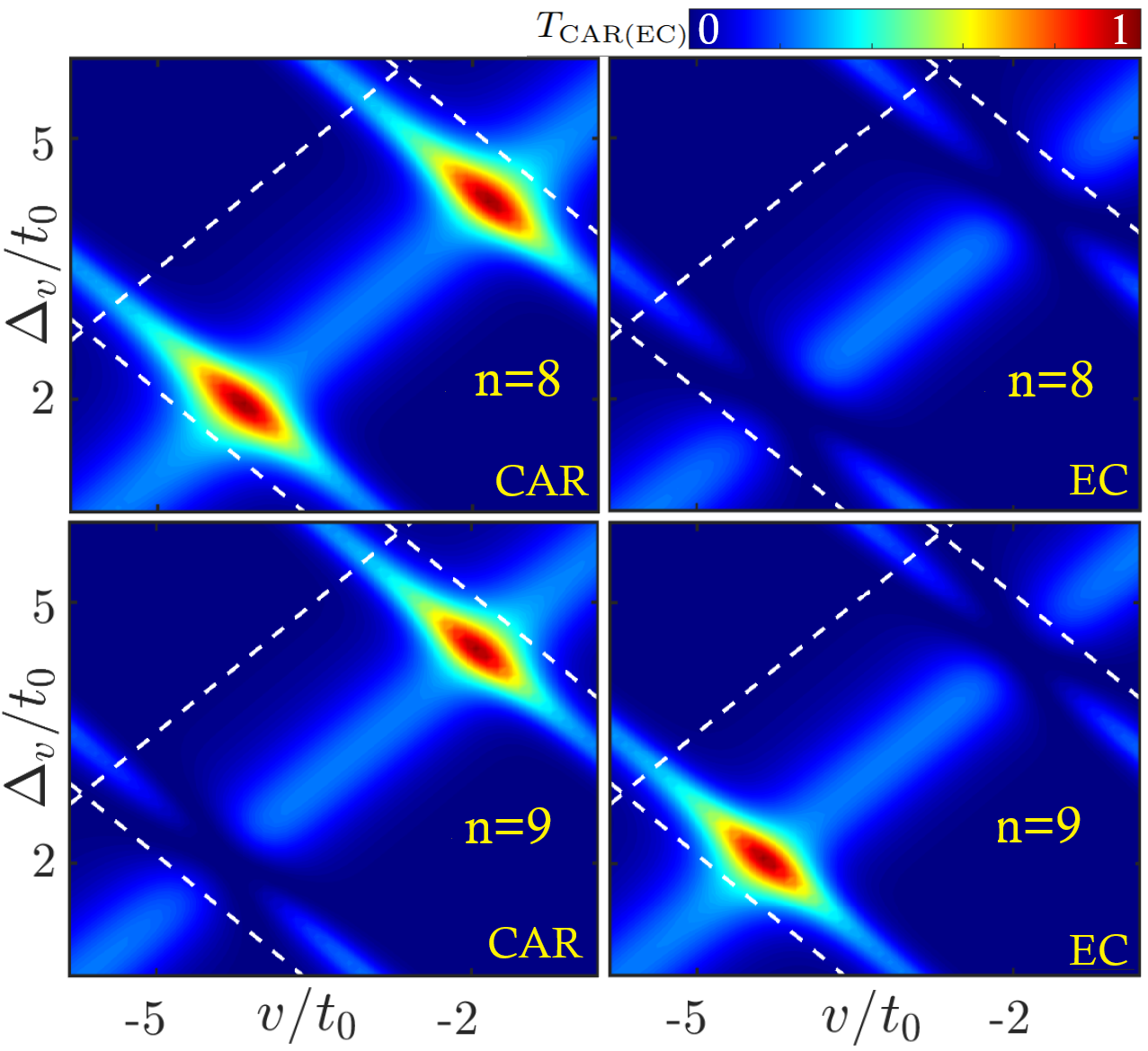}
\caption{\Gls{car} and \gls{ec} probabilities for the nonlocal conductance shown in \cref{fig2}c) with $n=8$, and \cref{fig2}d) with $n=9$.
}\label{fig4}
\end{figure}

\emph{Transmission for decoupled effective chains.}---
The zero-bias nonlocal conductance for the case of decoupled chains ($\mu_{A,B}=0$), shown in \cref{fig2}c) for $n=8$ and \cref{fig2}d) for $n=9$, is only finite around points $P$ and $Q$. However, the corresponding nonlocal transmissions $T_\text{CAR,EC}$ are finite for longer regions around the borders of the topological regions, marked by dot-dashed gray lines in \cref{fig2}c,d). We plot the transmissions in \cref{fig4}. While \gls{car} and \gls{ec} probabilities are very different at crossing points $P$ and $Q$, as described in the main text, they become the same at the boundaries away from the crossings. 
These areas correspond to peaks where only one of the effective chains contributes to the transmission. 
Therefore, according to \cref{eq:car-ec-red}, the amplitudes satisfy $T_\text{CAR}=T_\text{EC}$ and, consequently, they cancel out leading to a vanishing nonlocal conductance in \cref{fig2}. 
A similar behavior is observed in the system with finite onsite energies, as seen in the \gls{car} and \gls{ec} maps of \cref{fig3}. In the regions with intermediate intensity, only one of the effective chains contributes to the transport.

\begin{figure}[h]
\includegraphics[keepaspectratio=true,width=1.0\linewidth
]{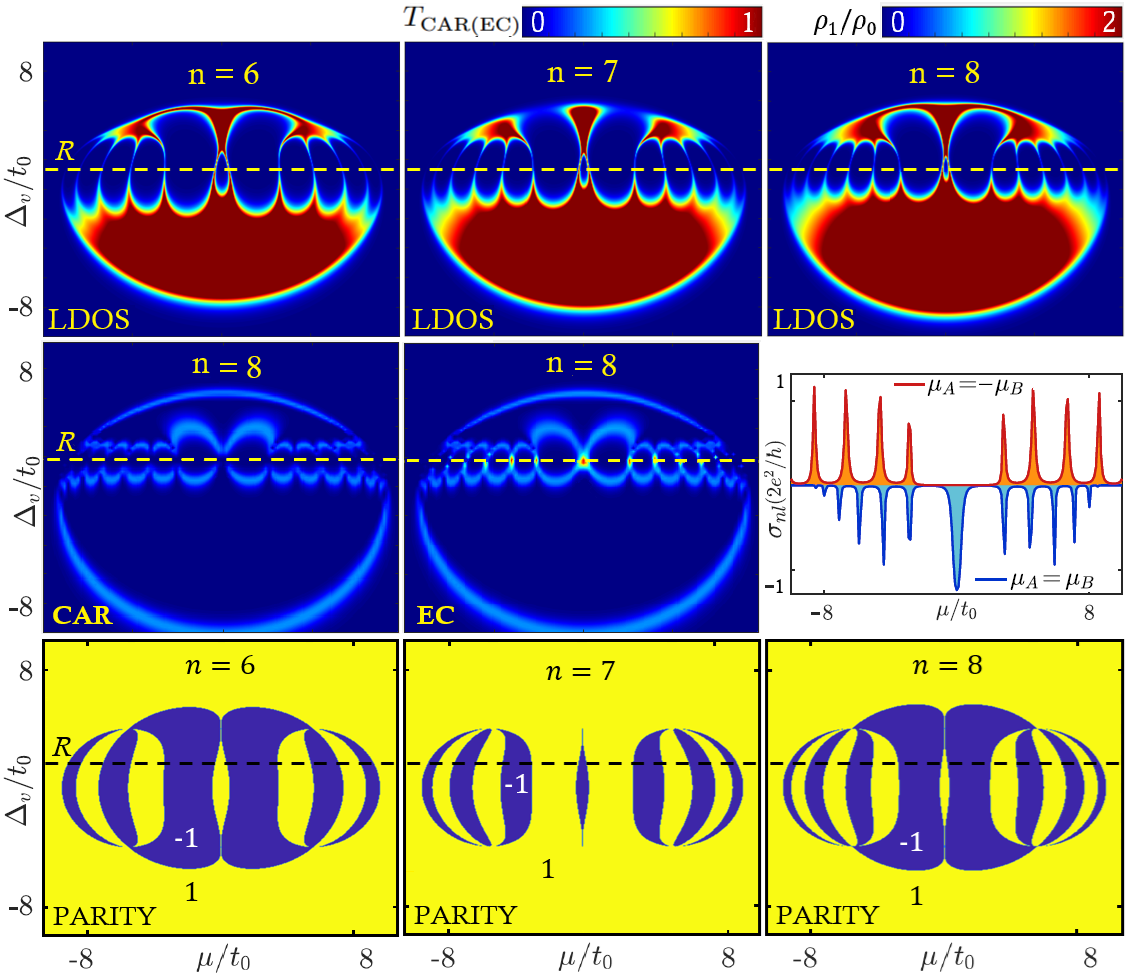}
\caption{\Gls{dos} (top), $T_\text{CAR,EC}$ (middle), and parity (bottom) as a function of $\Delta_v$ and $\mu\equiv \mu_A=\mu_B$. 
We set $\Delta_{w}=0.3$ and $w=0.3$ ($\Delta^{-}_{w}<0$).
For the \gls{dos} and parity we compare chains with 6, 7, and 8 unit cells. 
The nonlocal probabilities are computed for a chain with $n=8$, and the nonlocal conductance along line $R$ is shown on the right (blue lines), compared to the corresponding case for $\mu_A=-\mu_B$ (red lines).}
\label{fig5}
\end{figure}

\emph{Edge states and nonlocal transport for $\mu_{A}=\mu_{B}$.}---
In the main text we focused on the case with $\mu_A=-\mu_B$ which favors \gls{car} processes. For comparison we show in \cref{fig5} the case with $\mu_A = \mu_B$ and the same parameters. 
The \gls{dos} maps indicate that the topological phase diagram exhibits elliptic-shaped regions, consistently with \cref{eq:invariant}. The states near the line $R$ also decay slowly exhibiting an oscillatory behavior that extends along the entire chain. These oscillations lead to the resonance peaks at points where the finite-size parity changes sign. These features reflect parity transitions in the nonlocal transport response and align with the interference patterns and spatial distribution of the edge states.

Along line $R$, and in contrast to the system with opposite onsite energies, constructive interference enhances \gls{ec} over \gls{car}, which is suppressed due to destructive interference. Therefore, the nonlocal conductance $\sigma_{nl}$ is negative (positive) when $T_{\mathrm{EC}} >T_{\mathrm{CAR}}$ ($T_{\mathrm{EC}} <T_{\mathrm{CAR}}$), corresponding to the condition $\mu_A = \mu_B$ ($\mu_A = -\mu_B$).

\textit{Spatial dependence of the \glspl{mbs} wavefunctions.}---
We expand the bulk Hamiltonian, \cref{eqH}, around the high-symmetry points $k=0,\pi$ using a $\mathbf k \cdot \mathbf p$ approximation, thus translating the lattice model into a continuum description in real space. The Hamiltonian thus becomes a differential operator acting on the wavefunction of the states in each chain $\Phi=(\phi_{\alpha\beta}, \phi_{\beta\alpha})^{T}$, namely, 
\begin{equation}
\left( 
\begin{array}{cc}
A_{1}\sigma _{1}+B_{1}\sigma _{2} & \chi \\ 
\chi & A_{2}\sigma _{1}+B_{2}\sigma _{2} 
\end{array}%
\right) \Phi
 =E\Phi,   
\end{equation}
with $A_{1,2}=\pm(\Delta^{\pm}_{v}+\eta\Delta^{\mp}_{w}-\eta\Delta^{\mp}_{w}\frac{a^{2}}{2}\partial^{2}_{x})$ and $B_{1,2}=\pm a\eta\Delta^{\mp}_{w}\partial_{x}$, $\eta=\pm 1$ corresponding to $k=0,\pi$ points, respectively. The matrix $\chi$ couples the effective chains and is defined as $\chi=\frac{1}{2}(\mu_{A}+\mu_{B})\sigma_{0}+\frac{1}{2}(\mu_{A}-\mu_{B})\sigma_{3}$. 

The resulting system of equations for $\mu
_{A}=\pm \mu _{B}$ is 
\begin{equation}\label{eq:sys-eqs}
(A_{2}A_{1}+B_{1}B_{2}\pm\frac{1}{2}\mu^{2})\sigma
_{0}\phi_{\alpha\beta} =i(B_{2}A_{1}-A_{2}B_{1})\sigma_{3}\phi _{\alpha\beta}.   
\end{equation}

The wavefunction $\phi_{\alpha\beta}$ is an eigenfunction of the Pauli matrix $\sigma _{3}$, i.e., $\sigma _{3}\phi _{\alpha\beta} =\eta_{1}\phi _{\alpha\beta}$, with $\eta_{1}=\pm 1$. 
The edge states have wavefunctions exponentially localized at the edges, $\Phi= (\phi_{1}, \phi_{2})^T \me^{\lambda x}$, with $\phi_{1(2)}\propto [1(0),0(1)]^T$ for $\eta_1= 1$ and $\phi_{1(2)}\propto [0(1),1(0)]^T$ for $\eta_1=-1$. 
Inserting these wavefunctions in \cref{eq:sys-eqs} we reach the secular equation for $\lambda$, 
\begin{equation}
\frac{1}{4}a^{4}\lambda^{4}+ \eta\frac{1}{2}a^{2}C^{+}\lambda^{2}-\eta aC^{-}\lambda +\frac{Q_{\eta}}{\Delta^{+}_{w}\Delta^{-}_{w}}%
 =0 ,
\end{equation}
with $C^{\pm}=\Delta _{v}^{-}(\Delta _{w}^{+})^{-1}\pm\Delta _{v}^{+}(\Delta _{w}^{-})^{-1}$. 

\begin{figure}[h]
\includegraphics[keepaspectratio=true,width=1\linewidth]{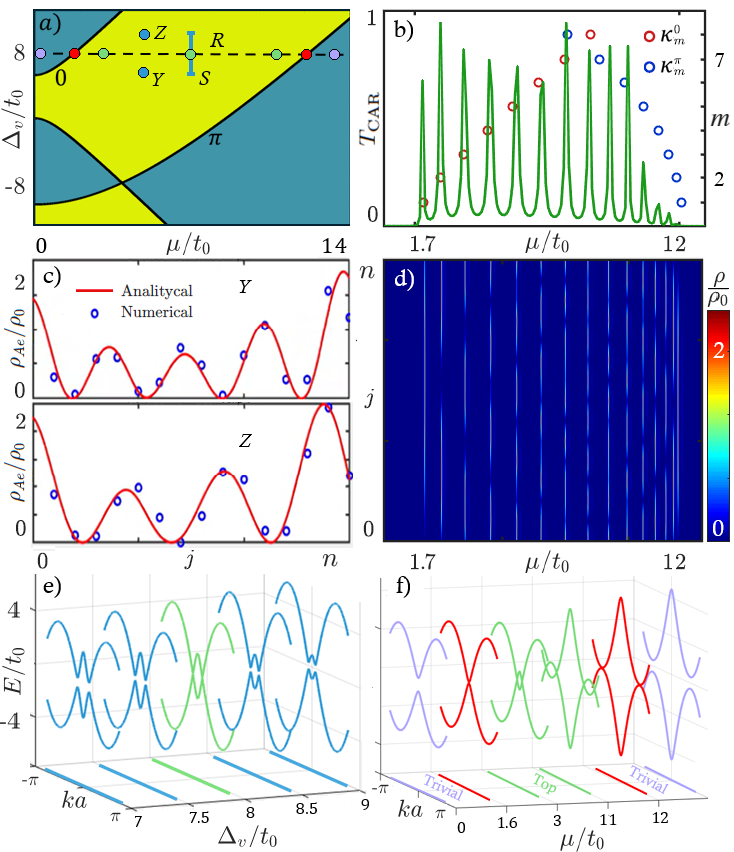}
\caption{a) Topological phase diagram of the system with $\mu_A=-\mu_B$ discussed in the main text near the $R$ line. The yellow regions correspond to configurations with nontrivial topology. 
b) $T_{CAR}$ vs $\mu$ at $E=0$ for $n=15$. The circles indicate the values of $\kappa$ at which a  mode $m$ emerges, corresponding to $\mu=\mu_m^{0,\pi}$. 
c) \gls{dos} (circles) at points Y (top, $m=5$ mode) and Z (bottom, $m=4$ mode) for electrons at $A$ atoms compared with the analytical results from the boundary wavefunction. 
d) \gls{dos} map for $n=15$ along the $R$ line. e) Energy band dispersion as a function of $k$, computed at different points along the vertical blue line in a). 
f) Energy band dispersion at the purple (trivial), red (gap closing), and green (topological) circles along line $R$ in a).
}
\label{fig6}
\end{figure}

The solutions for the decay parameter $\lambda$ can be either two real $\lambda_{1,2}= \xi_{1,2}^{-1}$ and two complex conjugate roots $\lambda^{\pm}_{3}=\xi_{3}^{-1} \pm i\kappa_{3}$, or two pairs of complex conjugate roots $\lambda^{\pm}_{1,2}=\xi_{1,2}^{-1} \pm i\kappa_{1,2}$ where the real part $\xi$ determines the decay length,  whose negative solution is for the state at the left edge of the chain, and the positive one for the state at the right edge. The $\kappa$ coefficient corresponds to the oscillation wavevector of the edge mode. In \cref{fig6}a) we present the topological phase diagram of the system with opposite onsite energies. Along line $R$ of the phase diagrams we have $C^{-}=0$, and the real part of the solutions vanishes. 
As a result, the edge states are completely delocalized $\xi\to\infty$, and $\kappa^{\eta}\approx a^{-1}\sqrt{|Q_{\eta}D_{\eta}^{-1}|}$, with $D_{\eta}=\eta\Delta^{+}_{w}\Delta^{-}_{w}C^{+}/2$, is proportional to the mass term $Q_{0,\pi}$ that governs the closing and reopening of the energy gap, cf. \cref{eq:invariant}. In \cref{fig6}b) we show $T_{CAR}$ along the $R$ line for $n=15$. Each maximum appears when a normal mode $m$ emerges for $an\kappa_{m}^{\eta}=m\pi$, namely, at
\begin{equation}
    \mu^{\eta}_{m}\approx\sqrt{D_{\eta}C^{+}\pi ^{2}\left( m/n\right) ^{2}+\left( \Delta
_{v}^{+}+\eta \Delta _{w}^{-}\right) \left( \Delta _{v}^{-}+\eta \Delta
_{w}^{+}\right) }. 
\end{equation}
In \cref{fig6}c), we illustrate the \gls{dos} at the $Y$ and $Z$ points near the $R$ line where the states decay exponentially. We compare with the results derived from the analytical approach $\rho \propto\left[\cos(\kappa_{m}^0 x+\phi)(e^{- x/\xi }+be^{x/\xi })\right]^2$. The \gls{dos} at points $Y$ and $Z$ corresponds to a normal mode with values $m=5$ and 4 respectively. In each of these normal modes, an extended state is present and the fermion parity is degenerate.  In \cref{fig6}d), the \gls{dos} map along the $R$ line is illustrated for different values of the chemical potential, where we can observe that when $\mu=\mu_m^{0(\pi)}$ a normal mode is formed.  At the $R$ line the condition $C^{-}=0$ is equivalent to 
\begin{equation}
    \frac{\Delta_{v}^{-}}{\Delta_{w}^{+}}=\frac{\Delta_{v}^{+}}{\Delta_{w}^{-}}. 
\end{equation}
Therefore, the $\alpha\beta$ chain is topologically equivalent to the $\beta\alpha$ chain, as both exhibit an identical ratio between intracell ($v$) and intercell ($w$) hopping amplitudes. This condition leads to a double crossing of the bands. 

\cref{fig6}e) displays the energy spectrum around the $R$ line, for the blue vertical $S$ line in \cref{fig6}a), where we observe that the gap closes exactly along the $R$ line at values of $k$ that are not high-symmetry points. This gap closing is not associated with a topological phase transition, but it leads to degenerate states that are extended along the chain and oscillate with wave vector $\kappa^{0,\pi}$. 

In \cref{fig6}f), we plot the energy spectrum at different points along the $R$ line, marked by colored circles in \cref{fig6}a), showing that the value of $k$ at which the gap closes varies continuously from $0$ to $\pi$, where the gap closes at $k = 0 (\pi)$ when $Q_{0(\pi)}$ changes sign.

\clearpage
\onecolumngrid
\setcounter{equation}{0}
\renewcommand{\theequation}{S\,\arabic{equation}}
\setcounter{figure}{0}
\renewcommand{\thefigure}{S\,\arabic{figure}}

\section{Supplemental Material to ``Topological superconductivity in a dimerized Kitaev chain revealed by nonlocal transport''}

In this section, we describe the Bogoliubov-de Gennes Hamiltonian in the Bloch basis, from which we compute the energy spectrum and the $Z_2$ index. We then present the Green's function method, which we apply recursively to calculate the propagators of a finite chain. Based on these Green's functions, we derive expressions for the current and non-local conductance in terms of crossed Andreev reflection (CAR) and elastic cotunneling (EC). Finally, we introduce the method used to compute the fermionic parity of the system.

\subsection{Energy bands and topological invariant\label{sec:app:Bulk}} 

\emph{Bulk system.}---
Imposing periodic boundary conditions to the Hamiltonian in \cref{eqH} we reach the bulk Bogoliubov-de Gennes Hamiltonian 
\begin{equation}\label{eq:app:H}
H(k)= 
\begin{pmatrix}
\mu _{A} & v+we^{ika} & 0 & \Delta _{v}+\Delta _{w}e^{ika} \\ 
v+we^{-ika} & \mu _{B} & -\Delta _{v}-\Delta _{w}e^{-ika} & 0 \\ 
0 & -\Delta _{v}-\Delta _{w}e^{ika} & -\mu _{A} & -v-we^{ika} \\ 
\Delta _{v}+\Delta _{w}e^{-ika} & 0 & -v-we^{-ika} & -\mu _{B}%
\end{pmatrix} ,
\end{equation}
with momentum $k$ and parameters defined in the main text. 
This Hamiltonian is written in the fermion basis $\Psi_{k}=(c_{Ak},c_{Bk},c^{\dagger}_{Ak},c^{\dagger}_{Bk})^{T}$. We can change into the Majorana representation by means of the unitary transformation $\Psi=U\bar{\gamma}$, with $\bar{\gamma}_{k}=(\alpha_{Ak},\beta_{Bk},\beta_{Ak},\alpha_{Bk})^{T}$ and
\begin{equation}\label{eq:U}
 U=\frac{1}{\sqrt{2}} 
\begin{pmatrix}
1 & 0 & i & 0 \\ 
0 & i & 0 & 1 \\ 
1 & 0& -i & 0 \\ 
0 & -i & 0 & 1%
\end{pmatrix} ,   
\end{equation}
where $U^{\dagger}U=\sigma_{0} \tau_0$. The resulting Hamiltonian takes the form
\begin{equation}
H_{\gamma }(k)= U^{\dagger}H(k)U = 
\begin{pmatrix}
0 & \Delta _{v}^{+}-\Delta _{w}^{-} \me^{iak} & \mu _{A} & 0 \\ 
\Delta _{v}^{+}-\Delta _{w}^{-} \me^{-iak} & 0 & 0 & \mu _{B} \\ 
\mu _{A} & 0 & 0 & -\Delta _{v}^{-}+\Delta _{w}^{+} \me^{iak} \\ 
0 & \mu _{B} & -\Delta _{v}^{-}+\Delta _{w}^{+} \me^{-iak} & 0%
\end{pmatrix} ,
\end{equation}
where the diagonal blocks 
\begin{equation}
    h_{\alpha\beta} = \left( \Delta_v^+ - \Delta_w^- \cos ka \right) \sigma_1 + \Delta_w^- \sin ka \sigma_2 
    , \quad 
    h_{\beta\alpha} = -\left( \Delta_v^- - \Delta_w^+ \cos ka \right) \sigma_1 - \Delta_w^+ \sin ka \sigma_2 
\end{equation}
correspond to the $\alpha \beta$ and $ \beta \alpha$ chains, which are coupled through the chemical potentials $\mu_{A,B}$. 
This Hamiltonian satisfies the Schr\"odinger equation $H_{\gamma}(k)\bar{\gamma}_{k}=E_k\bar{\gamma}$, with energy spectrum given by
\begin{equation}
    E_{k}=\pm \frac{1}{\sqrt{2}}\sqrt{|\sigma_{k}|^{2}+|\Lambda _{k}|^{2}+2\mu^{2}\pm \sqrt{Y_{k}}} ,
\end{equation}
where $Y_{k} =(|\sigma_{k}|^{2}-|\Lambda _{k}|^{2})^{2}+4\mu^{2}|\sigma_{k}-\Lambda _{k}|$, $\sigma_{k}=\Delta^{+}_{v}-\Delta^{-}_{w}e^{ika}$, and $\Lambda_{k}=\Delta^{-}_{v}-\Delta^{+}_{w}e^{ika}$. 

For decoupled chains ($\mu_{A}=\mu_{B}=0$) the spectrum reduces to 
\begin{align} \label{eq:gfgen}
E_{1k}={}&\pm |\Delta^{+}_{v}-\Delta^{-}_{w}e^{ika}| , \\
E_{2k}={}&\pm |\Delta^{-}_{v}-\Delta^{+}_{w}e^{ika}| .
\end{align}
This indicates that the cross-band condition for each effective chain retains a form similar to that of the conventional dimerizable \gls{ssh} model, where the intercell coupling equals the intracell coupling. That is, for the $\alpha\beta$ chain the condition is $\Delta^{+}_{v}=\eta\Delta^{-}_{w}$, while for the $\beta\alpha$ chain we have $\Delta^{-}_{v}=\eta\Delta^{+}_{w}$. 
The term $\eta=\pm 1$ corresponds to the point in the Brillouin zone where the cross-band process occurs ($k=0,\pi$). 
The phase diagram for the uncoupled system results in the set of dashed straight lines defined in \cref{fig1} of the main text. 

\emph{Topological invariant.}---
To define a topological invariant we transform \cref{eq:app:H} into the matrix $H_{s}(k)=U_{s}^{\dagger}H(k)U_{s}$, with
\begin{equation}\label{eq:U-trans}
    U_{s}=\frac{1}{2} 
\begin{pmatrix}
1 & i \\ 
1 & -i%
\end{pmatrix} \sigma_{0} ,
\end{equation}
defined through the set of Majorana operators such that $(c_{Ak},c_{Bk},c^{\dagger}_{Ak},c^{\dagger}_{Bk})^{T}=U_{s}(\alpha_{Ak},\alpha_{Bk}, \beta_{Ak},\beta_{Bk})^{T}$ which differs from $U$, the matrix that maps to the effective-chain basis. 
The resulting matrix $H_s$ is skew-symmetric, $H_{s}^{T}(k_s)=-H_{s}(k_s)$, when evaluated at the high-symmetry points $k_s=0,\pi$, and the Pfaffian at these points is linked to the product of the eigenenergies of \cref{eq:app:H}, i.e., $\mathrm{Pf}[H_{s}(0)] \mathrm{Pf}[H_{s}(\pi)]=\prod^{n}_{j=1}E_{j}$. Therefore, the Pfaffian changes sign whenever one of the occupied bands is inverted, which allows us to define a $\mathbb{Z}_2$ index 
\begin{equation}\label{eq:invariant1}
(-1)^\nu= \prod_{k=0,\pi}\mathrm{sgn} \{ \mathrm{Pf} [H_s(k)] \} = \mathrm{sgn}(Q_{0})\mathrm{sgn}(Q_{\pi}). 
\end{equation}

Specifically, we have 
\begin{equation} \label{eq1}
    H_{s}(k=0,\pi)= 
    \begin{pmatrix}
    0 & X \\ 
    X^{\dagger} & 0%
    \end{pmatrix} ,
\end{equation}
with
\begin{equation} \label{eq1}
X=\left( 
\begin{array}{cc}
\mu_{A} & \Delta^{+}_{v}+\eta\Delta^{-}_{w} \\ 
-\Delta^{-}_{v}+\eta\Delta^{+}_{w} & \mu_{B}%
\end{array}%
\right), 
\end{equation}
so that
\begin{equation}
Q_{\eta}=\left( \Delta _{v}^{+}+\eta\Delta
_{w}^{-}\right) \left( \Delta _{v}^{-}+\eta\Delta _{w}^{+}\right) +\mu_{A}\mu_{B}. 
\end{equation}

Thus, if the invariant takes the value $+1$, the system is in a topological phase, while a value of $-1$ corresponds to a trivial phase, as shown in the phase diagram in \cref{fig1}d). The parameter-dependent functions $Q_{\eta}$ indicate the topological phase transition curves where a gap closing occurs in the energy bands, and can be obtained directly from the energy spectrum by evaluating them at $E=0$ [solid red lines in \cref{fig1}d) for the coupled case, dashed white lines for the uncoupled case]. 

\subsection{Green functions\label{sec:app:GF}}

\textit{Recursive Green functions for finite chains.}---
To obtain the LDOS of each cell of the finite chain, $\rho_i = -\mathrm{Im}/[\mathrm{Tr}(\hat{G}_{ii})]/\pi$, we use the recursive GF coupling approach. 

\begin{equation}
\begin{aligned}
H_{0} &=
\begin{pmatrix}
\mu_{A} & v & 0 & \Delta_{v} \\
v & \mu_{B} & -\Delta_{v} & 0 \\
0 & -\Delta_{v} & \mu_{A} & -v \\
\Delta_{v} & 0 & -v & \mu_{B}
\end{pmatrix}
\qquad
\Sigma_{LR}=\Sigma^{*}_{RL} &=
\begin{pmatrix}
0 & w & 0 & \Delta_{w} \\
0 & 0 & 0 & 0 \\
0 & -\Delta^{*}_{w} & 0 & -w \\
0 & 0 & 0 & 0
\end{pmatrix}
\end{aligned}
\end{equation}

Starting from a single cell Green function $\hat{g}_{jj}=((E+i0^{+})\mathbb{I}-H_{0})^{-1}$, a chain with $n$ sites is obtained by iteratively coupling single sites via Dyson’s equation with hopping $\hat{\Sigma}_{ij}$. 
As a result, the local Green function at a specific cell $j$ is given by
\begin{equation}
\hat{G}_{jj}=\hat{g}_{jj}+\hat{g}_{jk}\hat{\Sigma}_{k,k'} \left[ \mathbb{I} -\hat{g}_{k'k'}\hat{\Sigma}_{k'k} \hat{g}_{kk}\hat{\Sigma}^{T}_{kk'} \right]^{-1} \hat{g}_{k'k'}\hat{\Sigma}^{T}_{k'k} \hat{g}_{kj} ,
\label{eq5}
\end{equation}
with $\mathbb{I}$ the identity matrix.

The self-energy $\hat{\Sigma}_{ij}$ couples a left subchain (cells 1 to $i$) with a right subchain (cells $j$ to $n$) with their edge GF $g_{ii}$ and $\hat{g}_{jj}$ respectively. These local functions of the chain edges can be computed recursively using this equation, if in \cref{eq5} is  $g_{ii}=g_{11}$ the GF of one cell, and substituting the subsequent coupling $g_{22}\rightarrow G_{11}$.
The non-local GF $\hat{g}_{1,n}$, which describes the propagation amplitude between chain edges, is computed with a perturbation at an intermediate cell $l$. At condition $i < l < j$, the Dyson equation takes the form 
\begin{subequations}\label{eq:car-ec_2}
\begin{align}
\hat{G}_{ij}=\hat{g}_{il}\hat{\Sigma}_{l,l'}\hat{G}_{l',j}\\ 
\hat{G}_{l',j}=\hat{g}_{l',j}+\hat{g}_{l',l'}\hat{\Sigma}_{l',l}\hat{G}_{l,j}\\
\hat{G}_{l,j}=\hat{g}_{l,l}\hat{\Sigma}_{l,l'}\hat{G}_{l',j}
\end{align}
\end{subequations}
replacing (c) in (b) and then in (a) 
\begin{equation}
    \hat{G}_{ij}=\hat{g}_{i,l}\hat{\Sigma}_{ll'} (\hat{1}-\hat{g}_{l'l'}\hat{\Sigma}_{l'l}
\hat{g}_{l,l}\hat{\Sigma}^{T}_{ll'}
)^{-1}\hat{g}_{l'j}
\end{equation}
By contrast, if $i < j < l$, the function assumes a different form
\begin{subequations}\label{eq:car-ec_3}
\begin{align}
\hat{G}_{ij}=\hat{g}_{ij}+\hat{g}_{il}\hat{\Sigma}_{l,l'}\hat{G}_{l',j}\\ 
\hat{G}_{l',j}=\hat{g}_{l',l'}\hat{\Sigma}_{l',l}\hat{G}_{l,j}\\
\hat{G}_{l,j}=\hat{g}_{l,j}+\hat{g}_{l,l}\hat{\Sigma}_{l,l'}\hat{G}_{l',j}
\end{align}
\end{subequations}
replacing (c) in (b) and then in (a)    
\begin{equation}
    \hat{G}_{ij}=\hat{g}_{i,l}\hat{\Sigma}_{ll'} (\hat{1}-\hat{g}_{l'l'}\hat{\Sigma}_{l'l}
\hat{g}_{l,l}\hat{\Sigma}^{T}_{ll'}
)^{-1}\hat{g}_{l'l'}\hat{\Sigma}_{l'l}\hat{g}_{l,j}.
\end{equation}

These functions can be generalized recursively to add more unit cells and model the propagator between the edge points of t   he structure, which connect the electrodes and allow the study of transport phenomena.

\subsection{Differential conductance\label{sec:app:cond}}

To study of nonlocal transport, we coupe semi-infinite metallic electrodes at each edge of the finite dimerized Kitaev chain. Within Keldysh formalism, we incorporate the effect of such leads via appropriate self-energy terms: $\hat\Sigma_{L,1}$ couples the left electrode to the sublattice atom $A$ of leftmost cell in the chain, while $\hat\Sigma_{R,n}$ couples the right electrode to the sublattice atom $B$ of the opposite edge, with

\[
\hat\Sigma_{L,1}\equiv \hat\Sigma_{L}=t_{L}\begin{pmatrix}
0 & 1 \\
0 & 0
\end{pmatrix} \tau_3, 
\quad
\hat\Sigma_{R,n} \equiv \hat\Sigma_{R} =t_{R}\begin{pmatrix}
0 & 0 \\
1 & 0
\end{pmatrix} \tau_3 .
\]
Here, $t_{L(R)}$ are the hopping amplitudes into the metallic leads and the Pauli matrices $\tau_{0,1,2,3}$ cat in Nambu (electron-hole) space. 

The average current in the left electrode, under an applied bias $V_{R}$ on the right electrode, is defined as the sum of electron–electron and electron-hole transmission processes. It is computed using the nonlocal Green function $\hat{G}^{1n}$ as

\begin{align}
    I^{e}_{L}(V_{R}) ={}& \frac{2e}{\hbar}
    \int
    \mathrm{Tr}[\mathrm{Re}(
    \hat{\Sigma}^{*}_{L}\hat{\rho}_{L}\hat{\Sigma}_{L}(E)\hat{G}_{eh}^{1n}(E)\hat{\Sigma}^{*}_{R}\hat{\rho}_{R}(E-eV_{R})\hat{\Sigma}_{R}\hat{G}_{he}^{1n}(E))]  
    [f(E-eV_{R})-f(E)] \md E,
    \\
    I^{h}_{L}(V_{R}) ={}& \frac{2e}{\hbar}
    \int
    \mathrm{Tr}[\mathrm{Re}( 
    \hat{\Sigma}^{*}_{L}\hat{\rho}_{L}(E)\hat{\Sigma}_{L}\hat{G}_{ee}^{1n}(E)\hat{\Sigma}^{*}_{R}\hat{\rho}_{R}(E-eV_{R})\hat{\Sigma}_{R}\hat{G}_{ee}^{1n}(E))]  
    [f(E-eV_{R})-f(E)] \md E,
\end{align}  
with $f(E)=[1+\exp(E/(k_BT))]^{-1}$ the Fermi distribution function at temperature $T$ ($k_B$ being the Boltzmann constant). 

\begin{figure}[ht]
\begin{center}
\advance\leftskip-3cm
\advance\rightskip-3cm
\includegraphics[keepaspectratio=true,scale=0.35]{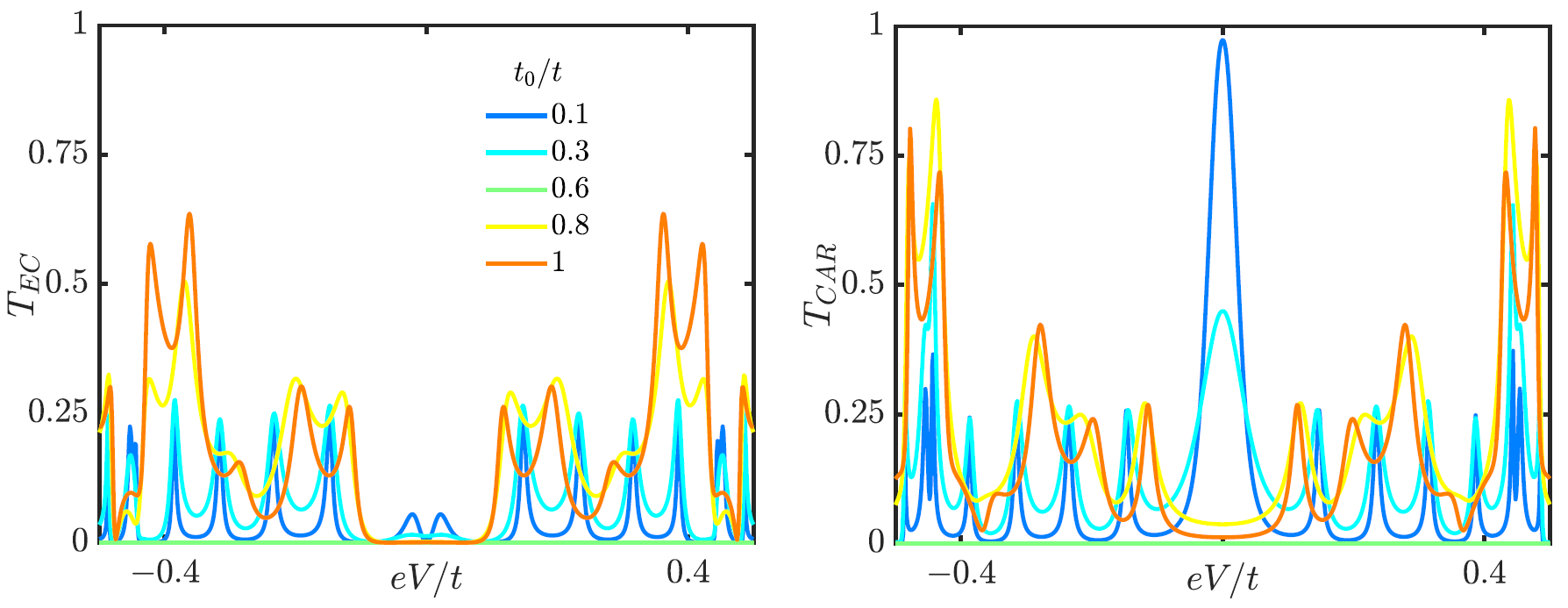}
\caption{\Gls{ec} (left) and \gls{car} (right) transmission as a function of bias voltage for the parameter set corresponding to point $P$ in \cref{fig2}c).
}
\label{figSM1}
\end{center}
\end{figure}

The differential conductance is the derivative of the current with respect to the bias voltage at zero temperature 
\begin{equation}
    \sigma_{nl}= \frac{d I_{L} }{d V_{R}}  = \frac{2e^{2}}{h}(T_\text{CAR}-T_\text{EC}), 
\end{equation}
with the scattering probabilities defined as
\begin{subequations}\label{eq:car-ec_5}
\begin{align}
T_{EC}(V)={}& \mathrm{Re}[\mathrm{Tr}(\hat{\Gamma}_{L,e} \hat{G}^{1n}_{ee} \hat{\Gamma}_{R,e} \hat{G}^{1n}_{ee})],\\
T_{CAR}(V)={}& \mathrm{Re}[\mathrm{Tr}(\hat{\Gamma}_{L,e} \hat{G}^{1n}_{eh} \hat{\Gamma}_{R,h} \hat{G}^{1n}_{he})] ,
\end{align}
\end{subequations}
for self-energies $\hat{\Gamma}_{L(R)}=\hat{\Sigma}_{L(R)}\hat{\rho}_{L(R)}\hat{\Sigma}_{L(R)}$. 

The two transmissions that define the conductance are elastic cotunneling ($T_\text{EC}$), that describes coherent electron transmission between electrodes, and crossed Andreev reflections $T_\text{CAR}$, corresponding to nonlocal Cooper pair splitting processes where an electron enters from one electrode and a hole exits through the other. 

These processes become more transparent in the effective chain picture: a basis transformation $G_{\gamma}=U^{\dagger}G^{1n}_{AB}U$, with $U$ defined according to \cref{eq:U}, isolates the transport channels $\alpha\beta$ and $\beta\alpha$, allowing us to express the conductance matrix elements in terms of Green’s functions in the Majorana basis. The element contributing to \gls{car}, $\hat{G}^{1n}_{AB,eh}$, can be rewritten as
\begin{equation}
G_{eh}=G_{\alpha\beta}- G_{\beta\beta}-G_{ \alpha\alpha}+G_{\beta\alpha } , 
\end{equation}
with short-hand notation $\hat{G}^{1n}_{AB,\epsilon \epsilon^{\prime}} \equiv G_{\epsilon \epsilon^{\prime}}$. 
On the other hand, the term that contributes to \gls{ec} is $\hat{G}^{1n}_{AB,ee}$, which can be written as
\begin{equation}
G_{ee}=G_{\alpha\beta }+ G_{\beta\beta}-G_{\alpha \alpha }-G_{\beta\alpha} .
\end{equation}

The Majorana representation clarifies the behavior of the peaks observed in the \gls{car} and \gls{ec} conductance maps and can be interpreted as interference effects between the transmission resonances of each effective chain. 
When $\mu_{A} = p\mu_{B}$, the \glspl{gf} that couple the effective chains are $\hat{G}^{1n}_{ AB,\alpha\alpha}=p\hat{G}^{1n}_{ AB, \beta\beta}$, which allows us to recast the probabilities as
\begin{equation} \label{eq1}
\begin{split}
T_\text{CAR}= t^{4}_{0} \rho^{L}_{e}\rho^{R}_{h} |G_{\alpha \beta }+G_{\beta \alpha }-(1+p)G_{\beta\beta }|^{2} , \\
T_\text{EC}= t^{4}_{0} \rho^{L}_{e}\rho^{R}_{e} |G_{\alpha \beta}-G_{\beta \alpha}-(1-p)G_{\beta\beta }|^{2} .
\end{split}
\end{equation}
In \cref{figSM1} we show the voltage dependence of these probabilities for different values of the couplings $t_L=t_R\equiv t_0$; we set $t_0=0.1t$ in the main text.



\subsection{Charge parity of a finite chain\label{sec:app:parity}}

To compute the charge parity of a finite chain and distinguish between even and odd chains, we write the Hamiltonian $H$ in \cref{eqH} with $n$ unit cells in the BdG form using the Nambu basis $\Psi =(c_{1A},c_{1B},\cdots ,c_{nA},c_{nB},c_{1A}^{\dag
},c_{1B}^{\dag },\cdots ,c_{nA}^{\dag },c_{nB}^{\dag })$. We then perform a rotation by defining a new basis vector, similar to the one used for computing the Pfaffian, but containing the Majorana operators of all atoms in the chain. This new basis is denoted as $%
\gamma =(\alpha _{1A},\alpha _{1B},\cdots ,\alpha _{nA},\alpha _{nB},\beta
_{1A},\beta _{1B},\cdots ,\beta _{nA},\beta _{nB})$, and the transformation reads $\Psi=\bar{U}_{s}\gamma$, with $\bar{U}_s = U_s \otimes \mathbb{I}_{n\times n} $ and $U_s$ defined in \cref{eq:U-trans}. 
\begin{equation}
H= \Psi^{\dagger} H_{BdG }\Psi = \gamma^{\dagger} \bar{U}_{s}^{\dagger}H_{BdG} \bar{U}_{s}\gamma = \gamma^{\dagger} H_{s}\gamma ,
\end{equation}
where $H_s = \bar{U}_s^\dagger H_{BdG} \bar{U}_s$ a skew-symmetric matrix. 
Its eigenvalues come in purely imaginary conjugate pairs $\pm iE_{jA,B}$ with $j=1,2..., n$ and satisfies the eigenvalue equation
\begin{equation}
H_{s}v_{i}=E_{i}v_{i} .
\end{equation}
Since $H_{s}$ has only real-valued entries, its eigenfunctions come in pairs as $\pm v_{i}$ for $\pm E_{i}$ eigenvalues. With this vectors we define the matrix $O=[v_{1A},v^{*}_{1A}, v_{1B}...v_{nB},v^{*}_{nB}]$ that satisfies the matrix equation
\begin{equation}
H_{s}O=OD, 
\end{equation}
with the diagonal matrix $D=\mathrm{diag}\{ -iE_1,iE_1, \dots, -iE_n,iE_n \}$. 
The matrix $O$ transforms $H$ into a block-diagonal form, allowing the Pfaffian to be expressed as the product of self-energies
\begin{equation}
 O^{T}H_{s}O=i
\begin{pmatrix}
0 & -E_{1} & 0 & 0 & \cdots & 0 & 0 \\
E_{1} & 0 & 0 & 0 & \cdots & 0 & 0 \\

0 & 0 & 0 & -E_{2} &                & 0 & 0 \\
0 & 0 & E_{2} & 0 &                & 0 & 0 \\

\vdots & \vdots &        &         & \ddots & \vdots & \vdots \\

0 & 0 & 0 & 0 & \cdots  & 0 & -E_{n} \\
0 & 0 & 0 & 0 & \cdots  & E_{n} & 0
\end{pmatrix}.
\end{equation}
Its determinant, $P = \det(O) = \pm 1$, encodes the ground-state charge parity of the system, capturing whether the system hosts an even or odd number of fermions. In \cref{fig3}a) we show that the constructive and destructive interference peaks arise from sign changes in the charge parity.

\end{document}
%